Research Article

# Review of Machine Learning Techniques for Power Electronics Control and Optimization

Maryam Bahrami[1], Zeyad Khashroum[2]*

[1] Department of Industrial Engineering, Lamar University, Beaumont, TX 77710, USA
[2] Department of Electrical Engineering, Lamar University, Beaumont, TX 77710, USA

| Keywords | Abstract |
|---|---|
|  | In the rapidly advancing landscape of contemporary technology, power electronics assume a pivotal role across diverse applications, ranging from renewable energy systems to electric vehicles and consumer electronics. The efficacy and precision of these power electronics systems stand as cornerstones of their functionality. Within this context, the integration of machine learning techniques assumes paramount significance. This article endeavors to present an extensive and comprehensive review of the machine learning techniques that find application in power electronics control and optimization. Through meticulous exploration, we aim to elucidate the profound potential of these methods in shaping the future of power electronics control and optimization. |

## 1. Introduction

Power electronics refers to the application of solid-state electronics to control and convert electrical power. These systems are found in various domains, including industrial automation, renewable energy, transportation, and more. The efficiency and reliability of power electronic systems are vital for their successful operation [1–8].

Machine learning techniques have been applied to power electronics control and optimization to improve the performance of power electronics systems [9–11]. These techniques can be used to reduce the computational expense associated with characterizing DC-DC converters, which is necessary for designing and optimizing power electronics systems. Machine learning techniques such as support vector regression and artificial neural networks have been utilized to accurately predict DC-DC converters' performance [12–17].

In addition, several publications have reviewed artificial intelligence (AI) applications for power electronic systems [7,18–20]. These applications include optimization, classification, regression, and data structure exploration and can be applied to the design, control, and maintenance phases of the power electronics system lifecycle [10]. AI techniques such as expert systems, fuzzy logic, metaheuristic methods, and machine learning have been discussed in the literature [10,21].

The existing application of machine learning methods for enhancing power system resilience has been reviewed in [20], which is the ability to withstand and recover from extreme events or attacks. The paper discusses the challenges and opportunities of applying machine learning for various aspects of power system resilience, such as situational awareness, contingency analysis, restoration planning, and adaptive protection. The paper also describes machine learning techniques and their applications in power system resilience.

Modeling Energy Consumption Using Machine Learning has been introduced in [22]. This paper develops predictive models for energy consumption using machine learning techniques. The paper compares the performance of different machine learning techniques, such as Multiple Linear Regression, Random Forest Regressor, Decision Tree Regressor, Extreme Gradient Boost Regressor, Support Vector Machines, K-Nearest Neighbor, and deep learning. It has been shown that Random Forest Regressor and deep learning have the best accuracy among the tested techniques.

In [23], a systematic review of the fault detection and diagnosis techniques for complex systems and technologies has been presented. Fault detection and diagnosis are processes that monitor the system health and identify the causes and locations of malfunctions. The paper categorizes the techniques into model-based and data-driven approaches, focusing on artificial intelligence-based methods. The article also describes the typical steps involved in designing and developing automatic fault detection and diagnosis systems.

In another study [24], analysis of power flows, power quality, photovoltaic systems, intelligent transportation, and load forecasting have been discussed in detail. The survey investigates the most recent and promising ML techniques proposed by the literature, highlighting their main characteristics and relevant results. The review revealed that ML algorithms can handle massive quantities of data with high dimensionality, allowing the identification of hidden characteristics of complex systems. Hybrid models generally show better performances when compared to single ML-based models.

Solar power generation can be unpredictable due to cloud cover and weather conditions [25]. This study uses meteorological data to predict solar power output, with Support Vector Regression proving to be a more effective method than other machine learning algorithms [25]. Various parameter tuning techniques, including Random search, Grid search, and Tree-based optimization, are employed to create a robust model for accurate solar power prediction.

Traditional power electronics systems for fuel cell-powered electric vehicles often involve two separate boost converters, which can be challenging for high-density vehicle applications. In [26], the authors discussed methods for designing an optimal system that integrates fuel cells and batteries in electric vehicles. It proposes a solution using reinforcement learning to adapt to changing conditions, resulting in higher efficiency and reduced carbon emissions, which is especially beneficial for heavy-duty commercial vehicles.

The field of power electronics, machine drives, and electric vehicles increasingly focuses on data-driven fault classification for power converter systems. A data-driven, supervised machine learning approach that combines Expectation Maximization Principal Component Analysis (EMPCA) and Support Vector Machine (SVM) to classify different fault topologies in real-time control systems has been discussed in [27]. The methodology is tested on non-inverting Buck-Boost DC-DC power converters for various fault scenarios, demonstrating its feasibility through intensive simulations and comparison studies.

Overall, machine learning techniques which are depicted in Figure 1 offer a promising approach for improving the performance of power electronics systems through control and optimization [28,29]. Further research is needed to fully realize the potential of these techniques in this field. In this review, we discuss the role of control and optimization, and traditional approaches. We will also discuss SVM and neural networks, and their impact on power electronics. Figure 2 shows some real-world applications of ML algorithms.

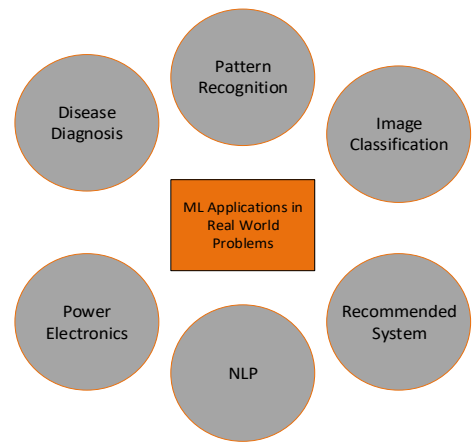

**Figure 1.** Various machine learning algorithms

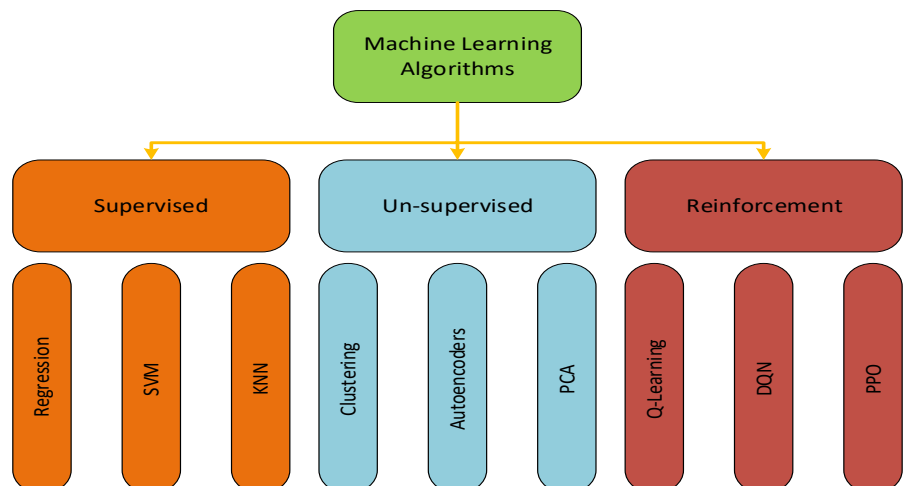

**Figure 2.** Real world applications of machine learning

## 2. Methodology of the Review

The main goal of this systematic review is to offer an in-depth analysis of the machine learning techniques for power electronics control and optimization. The following are the most important research concerns this work addresses:

- Advancements in power electronics technology
- Efficacy and precision
- Integration of machine learning

For this study, a sizable amount of literature was carefully screened in order to gather pertinent research articles. To this extent, peer-reviewed journal articles, conference papers, research pieces, and review articles were gathered from all popular databases including Science Direct, Google Scholar, Scopus, IEEE, etc. The fields of power electronics control and optimization, such as power electronics converter, control strategies, and power electronics applications, were where the majority of the research-gathering process's keywords were targeted.

## 3. The role of Control and Optimization

Control and optimization play a crucial role in power electronics systems. Power electronics systems are used to convert and control electrical energy, and they are found in a wide range of applications, from renewable energy systems to electric vehicles. Control algorithms are used to regulate the output of these systems, ensuring that they operate efficiently and reliably. Optimization techniques, on the other hand, are used to improve the performance of power electronics systems by finding the best operating conditions [9–11].

In recent years, machine learning techniques have been applied to power electronics control and optimization to improve the performance of these systems. These techniques can be used to reduce the computational expense associated with the characterization of dc-dc converters, which is necessary for the design and optimization of power electronics systems consisting of multiple converters. Machine learning techniques such as random forest and gradient boosting have been shown to be capable of accurately predicting the performance of commercially available dc-dc converters. Overall, control and optimization are essential for ensuring the efficient and reliable operation of power electronics systems [12–17].

## 4. Traditional Approaches vs. Machine Learning

Traditionally, control and optimization in power electronics relied on mathematical models and heuristic algorithms. Machine learning, however, offers a data-driven approach that can adapt to changing conditions and optimize performance in real-time. Traditional approaches for power electronics control and optimization involve using analog control techniques and sensor-based methods for temperature estimation [30]. These methods have been used for a long time and have proven to be effective. However, with the advent of machine learning (ML) and artificial intelligence (AI), there has been a shift towards using data-driven approaches for power electronics control and optimization [31].

ML techniques such as fuzzy logic, feed-forward neural networks, recurrent neural networks, and reinforcement learning are being developed for power electronics control and optimization [31]. These techniques allow for more complex and dynamic non-linear control surfaces to enhance efficiency, reliability predictions, and health monitoring in power converters.

As it is mentioned above, traditional approaches for power electronics control and optimization involve using analog control techniques and sensor-based methods, while machine learning-based data-driven approaches are becoming standard tools for the automated high-performance control and monitoring of electric drives [30].

## 5. Supervised Learning for Control

Supervised learning is a category of machine learning where an algorithm is trained using labeled data to understand the relationship between input and output variables [32]. In the context of power electronics control and optimization, supervised learning helps in forecasting how a system will behave based on input variables like voltage, current, and temperature [10,33–37]. Several supervised learning techniques are applicable in power electronics, including linear regression, support vector machines (SVMs), and neural networks [9,38–40].

### 5.1. Linear Regression

Linear regression is a straightforward and widely employed supervised learning method that handles both regression (predicting numerical values) and classification (assigning data points to categories) tasks. Its core principle is to determine the best-fit line that minimizes the sum of the squared differences between predicted and actual values. In the context of power electronics, linear regression can be used to predict the system's output voltage or current based on input parameters.

### 5.2. Support Vector Machines (SVMs)

SVMs are another prevalent supervised learning approach suitable for both classification and regression problems. SVMs function by identifying a hyperplane that effectively separates data into distinct categories while maximizing the margin between these categories. In power electronics control and optimization, SVMs can classify the system into different operational modes or predict output variables (like voltage or current) based on input parameters [38,41].

### 5.3. Neural Networks

Neural networks are more intricate supervised learning models designed to handle a broad array of tasks, including image recognition, speech processing, natural language understanding, and power electronics control and optimization.

They emulate the functioning of the human brain, comprising layers of interconnected nodes (neurons) that process and transmit data. In the power electronics context, neural networks can predict output variables (voltage or current) based on input parameters or optimize system parameters for specific applications.

As it is clear, supervised learning techniques like linear regression, support vector machines, and neural networks are valuable tools for power electronics control and optimization. They enable the prediction of system behavior based on input data and the fine-tuning of system parameters to achieve specific goals in power electronics applications [42–45].

## 6. Unsupervised Learning for Optimization

Unsupervised learning is a machine learning paradigm where the algorithm works without labeled data and aims to discover inherent patterns or structures within the data. In power electronics, unsupervised learning techniques play a significant role in optimization processes [46,47]. Clustering and Principal Component Analysis (PCA) discussed following are two key techniques of unsupervised learning [10,48–50].

### 6.1. Clustering

Clustering is a method used to group similar data points together based on their characteristics or features [47,51]. In the context of power electronics, clustering can be employed to the following.

### 6.1.1 Identify Patterns in Power Consumption Data

Clustering algorithms can group consumers or devices exhibiting similar power usage patterns. This information can be valuable for load balancing, as it helps utilities allocate resources more efficiently [52,53].

### 6.1.2 Detect Anomalies

Clustering can also be used to identify unusual or anomalous patterns in power consumption. For example, sudden spikes or drops in energy usage can indicate faults or irregularities in the power system. Detecting these anomalies is crucial for fault detection and preventive maintenance [54].

### 6.2. Principal Component Analysis (PCA)

PCA is a dimensionality reduction technique used to simplify complex datasets by transforming them into a lower-dimensional space [55]. In power electronics, PCA can be beneficial in the following ways:

### 6.2.1 Reducing Data Complexity

Power systems generate vast amounts of data. PCA helps by identifying the data's most significant components (principal components), thereby simplifying it [56].

### 6.2.2. More Efficient Optimization

Simplified data resulting from PCA can lead to more efficient optimization algorithms. These algorithms can be applied to load scheduling, power flow analysis, and system parameter tuning tasks.

As clearly described above, unsupervised learning techniques, such as clustering and Principal Component Analysis, are powerful tools in power electronics. Clustering helps identify patterns and anomalies in power consumption, aiding load balancing and fault detection. On the other hand, PCA reduces the dimensionality of complex power system data, making it more manageable for optimization processes ultimately contributing to more efficient power electronics systems.

## 7. Reinforcement Learning Application in Power Electronics

Reinforcement learning (RL) is a machine learning paradigm where an agent learns by interacting with an environment [20,57,58]. It aims to maximize a cumulative reward signal over time by making a sequence of decisions. In power electronics, RL offers unique advantages and applications as follows.

### 7.1 Adaptive Control

Power electronics systems often operate in dynamic and uncertain environments. RL suits these scenarios well because it allows systems to learn and adapt to changing conditions. RL algorithms can continuously adjust control parameters, such as voltage or current setpoints, to optimize system performance. This adaptability is particularly beneficial in systems with varying loads or renewable energy sources [59,60].

### 7.2. Energy Management

Energy management in power electronics involves optimizing the allocation and consumption of electrical energy. RL can be employed to make real-time decisions regarding energy distribution. RL algorithms can dynamically manage microgrids by deciding when to draw power from the grid, store energy, or use renewable sources based on real-time data and demand [61,62].

### 7.3. Fault Detection and Diagnosis

RL can play a role in the early detection and diagnosis of faults in power electronics systems. By learning the system's normal behavior, RL agents can identify deviations that may indicate defects or anomalies. Once a fault is detected, RL algorithms can suggest corrective actions or control adjustments to mitigate the impact of the fault, ensuring system reliability and safety.

### 7.4. Optimization of Renewable Energy

Renewable energy sources, like solar and wind, are inherently variable. RL can optimize the integration of these sources into the power grid by making real-time decisions on energy generation and distribution. RL can help balance the grid by adjusting the output of renewable sources, such as wind turbines, to maintain grid stability in fluctuations. As described above, reinforcement learning is a powerful tool for adaptive control and optimization in power electronics. Its ability to learn from environmental interactions makes it well-suited for managing complex and dynamic power systems, optimizing energy usage, detecting faults, and integrating renewable energy sources efficiently. As science advances, RL will likely play an increasingly crucial role in improving power electronics systems' reliability, efficiency, and sustainability [58,60].

## 8. Challenges in Implementing Machine Learning

Implementing machine learning in power electronics systems presents significant potential benefits, but it also comes with challenges related to data quality, model interpretability, and meeting real-time requirements. Overcoming these challenges is crucial to harness the full potential of machine learning for improving the efficiency, reliability, and performance of power electronics applications.

### 8.1. Data Collection and Preprocessing

High-quality data collection is the foundation of any successful machine-learning application [63]. In the context of power electronics, this means gathering accurate and comprehensive data related to voltage, current, temperature, and other relevant parameters. Data may come from various sources, such as sensors and monitoring devices placed within the power system. Ensuring the reliability and consistency of this data is critical.

Data preprocessing involves cleaning, transforming, and structuring the data before feeding it into machine learning models. This step is vital because raw data may contain noise, outliers, or missing values that could lead to inaccurate model predictions [64].

Data preprocessing also includes feature selection, where the most relevant variables are chosen to train the model effectively. Inaccurate or irrelevant features can negatively impact model performance.

### 8.2. Model Interpretability

In specific applications of power electronics, particularly those with safety or regulatory implications, it's essential to have interpretable machine learning models. Interpretable models are those whose decisions can be understood and explained in a human-readable manner. This is crucial for ensuring transparency and accountability. For instance, if a machine learning model is used to control voltage levels in a power grid, engineers and regulators need to know why the model made a particular decision in case of unexpected outcomes or errors [65].

Achieving model interpretability may require simpler algorithms or techniques that provide clear insights into the model's decision-making process. While complex deep learning models can offer excellent performance, they are often less interpretable than simpler models like decision trees or linear regression.

### 8.3. Real-time Constraints

Power electronics systems often operate in real-time environments, where decisions must be made rapidly to maintain stability and safety. Many machine learning algorithms are computationally intensive and may not be well-suited for real-time decision-making [66].

Meeting real-time constraints involves optimizing the machine learning model's efficiency and responsiveness. This can be achieved through techniques like model simplification, parallel processing, or using specialized hardware accelerators. Additionally, models may need to be continuously retrained or updated in real-time as new data becomes available to ensure they remain accurate and adaptive to changing conditions [67].

## 9. Discussion

In this study, we conducted an extensive literature review to ascertain the prevalence of various machine learning techniques in the realm of power electronics control and optimization. Our method involved comprehensive searches across reputable academic databases and specialized journals (for selected 24 papers in the literature), utilizing specific search queries tailored to each machine learning technique of interest, including linear regression, support vector machines, neural networks, clustering, and Principal Component Analysis (PCA). Subsequently, we meticulously filtered and analyzed the retrieved publications, focusing on their relevance to power electronics, control, optimization, and the specified machine learning methodologies. The obtained data enabled us to calculate the respective percentages of research works employing each technique. The resulting percentages, depicted in Figure 3, offer valuable insights into the relative adoption and utilization of these machine learning approaches within the field of power electronics control and optimization. As it is shown below, the percentages are as follows, linear regression (8 %), support vector machines (36 %), neural networks (22 %), clustering (28 %), and Principal Component Analysis (6 %). We have also compared several ML based models applied in power electronics control and optimization for better prediction. The comparison results and utilized metrices and parameters are also presented in Table 1.

- Linear regression
- Support vector machines
- Neural networks
- Clustering
- Principal Component Analysis

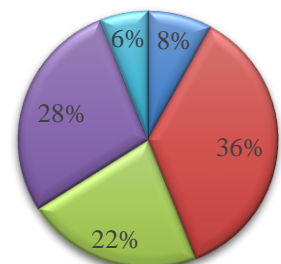

**Figure 3.** Percentage of machine learning algorithms utilized in the field of power electronics control and optimization.

Table 1. Comparison of various machine learning models

| References | Algorithms | Important parameters | Metrics | Values |
|------------|-----------|---------------------|---------|--------|
| [41] | SVM | Input voltage and switching frequency | $I_{rms}$ | 0.84 |
| [25] | SVM | Random Search | RMSE | 2254.66 |
| [68] | Random forest | Correlation coefficient | RMSE | 3.3714 |
| [69] | Neural network | A and λ | Overshoot (%) | 29 |
| [45] | 1-NN | TS COM | F1-score | >95% |
| [70] | NN | $K_p$ | Overshoot (%) | 0.976 |
| [71] | FF-NN | $V_{ref}$ | Voltage | 80 |
| [72] | NN | H, S, L, N, I | MAE | 0.85 |
| [73] | Auto-ML | DC link voltage | Voltage | 0.4 |
| [74] | KNN | $S_i$ | $R^2$ | 0.79 |

## 10. Conclusions

This article comprehensively overviews the symbiotic relationship between machine learning techniques and power electronics control and optimization. As we navigate the dynamic landscape of modern technology, it becomes increasingly apparent that power electronics are instrumental across a wide range of applications. The precision and efficiency of power electronics are non-negotiable facets of their functionality, and herein lies the significance of the integration of machine learning methodologies. Through this exploration, we have uncovered the vast potential of machine learning techniques in enhancing the control and optimization of power electronics systems. These methods empower us to make informed decisions, adapt to changing conditions, and achieve higher levels of energy efficiency. As we look ahead, it is evident that the synergy between machine learning and power electronics is poised to drive further innovation, revolutionizing how we harness and utilize electrical power in our interconnected world. The journey toward the future of power electronics control and optimization has begun, guided by the powerful beacon of machine learning.

## Conflict of Interest Statement

The authors declare no conflict of interest.